\documentclass[aps, pra, 10pt, twocolumn]{revtex4-2}
\usepackage[T1]{fontenc}
\usepackage[utf8]{inputenc}
\usepackage{units}
\usepackage{amsmath}
\usepackage{amssymb}
\usepackage{braket}
\usepackage{graphicx}
\usepackage{esint}
\usepackage{xcolor}
\definecolor{darkblue}{rgb}{0.0, 0.0, 0.75}
\usepackage{color}
\usepackage[colorlinks=true, linkcolor=darkblue, urlcolor=darkblue, citecolor=darkblue]{hyperref}

\begin{document}
\title{Self-correction phase transition in the dissipative toric code}
\author{Sanjeev Kumar}
\email{sanjeev.kumar@tu-berlin.de}
\author{Hendrik Weimer}
\affiliation{Institut f\"ur Physik und Astronomie, Technische Universit\"at  Berlin, Hardenbergstraße 36, EW 7-1, 10623 Berlin, Germany}
\date{\today}

\begin{abstract}
We analyze a time-continuous version of a cellular automaton decoder
for the toric code in the form of a Lindblad master equation. In this
setting, a self-correcting quantum memory becomes a thermodynamical
phase of the steady state, which manifests itself through the steady
state being topologically ordered. We compute the steady state phase
diagram, finding a competition between the error correction rate and
the update rate for the classical field of the cellular
automaton. Strikingly, we find that self-correction of errors is
possible even in situations where conventional quantum error
correction does not have a finite threshold.
\end{abstract}

\maketitle

\section{\label{sec:Introduction}Introduction}

The long-term preservation of quantum information is one of the key
requirements to build a scalable quantum computer
\cite{nielsen2010quantum, DiVincenzo2000, preskill1998reliable}. This had led to the development of
sophisticated error correction algorithms to combat unavoidable
imperfections during the operation of the device \cite{bausch2024learning, Delfosse2021almostlineartime, Baireuther2018machinelearning, terhal2015quantum, PhysRevA.90.032326MLD, hastings2013decoding, PhysRevA.86.032324, PhysRevLett.104.050504RG, gottesman1997stabilizer}. Here, we connect
these efforts to the emerging field of open quantum many-body systems \cite{Diehl2008,Verstraete2009,Weimer2010,Hartmann2010,Diehl2010a,Nagy2010,Tomadin2011,Lee2011,Kessler2012,Honing2012,Sieberer2013,Torre2013,Lee2013,Cai2013,Weimer2015,Marcuzzi2016,Maghrebi2016,Jin2016,Marino2016,Kshetrimayum2017,Rota2017,Buchhold2017,Nagy2018,Carollo2019,Panas2019,Raghunandan2020,Carollo2022,Singh2022,Mink2023,Sieberer2025} 
and show that a dissipative variant of the toric code exhibits a
self-correcting phase that can be used for long-term information
storage.

Quantum error correction algorithms are routinely characterized in
terms of their error correction threshold, which refers to the maximum
strength of a depolarizing noise channel applied to one of the logical
states that still can be corrected \cite{wang2003confinement}. While this assumes a static and
uncorrelated application of the noise, real-world quantum devices are
complex dynamical systems potentially exhibiting emergent errors that
cannot be understood purely from a microscopic set of rules \cite{Sarovar2020detectingcrosstalk, PhysRevLett.96.050504, wilen2021correlated}. Such a
viewpoint is much closer to the notions of open quantum many-body
systems, where the long-term behavior of quantum systems is
characterized in terms of thermodynamic phases of the steady state and
phase transitions between them \cite{Kumar2024, Vodola2022fundamental, minganti2018spectral, Kessler2012}.

In this article, we provide a synthesis between the fields of quantum
error correction and open quantum many-body systems. Specifically, we
consider a dissipative variant of the toric code model that is
governed by microscopic Lindblad dynamics \cite{roberts2020driven, bardyn2013topology}. The Lindblad jump operators
are structured in such a way that they implement a cellular automaton
decoder for the toric code \cite{Herold2015, herold2017cellular, kubica2019cellular, kubica2021cellular} and hence have the ground states of the
toric code as their dark states. In order to be able to fuse distant
anyon excitations by purely local jump operators, the rates for the
quantum jumps are controlled by the value of a classical field that is
updated according to the local anyon densities, in a similar way as it
has already been discussed for conventional error correction
algorithms \cite{harrington2004analysis, dennis2002topological}. To introduce errors, we consider the appearance
of local bit and phase flip errors; however, the subsequent evolution
naturally generates more complex error syndromes. We then
analyze the steady state of the dynamics and show that the dynamics is
capable of self-correction provided the state remains in the
topologically ordered phase according to a recently introduced
operational definition of topological order \cite{Jamadagni2022}. We map out the steady
state phase diagram by varying the relative strengths of the error
rates and field-update rates, respectively. Crucially, we find a
stable quantum memory even when the classical field is
two-dimensional, in contrast to the conventional error correction
scenario \cite{PhysRevA.109.022422, PhysRevA.91.062324, balasubramanian2025localautomaton2dtoric, dunnweber2026quantum}.

\section{\label{sec:Model}Dissipative toric code}

\subsection{The toric code model}

\begin{figure}[b]
  \begin{tabular}{cc}
    \includegraphics[height=3.5cm]{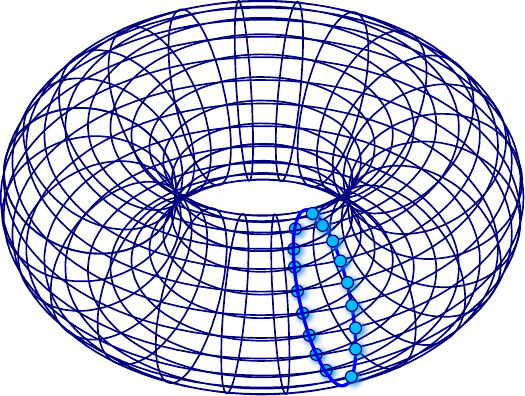} & \includegraphics[height=3.5cm]{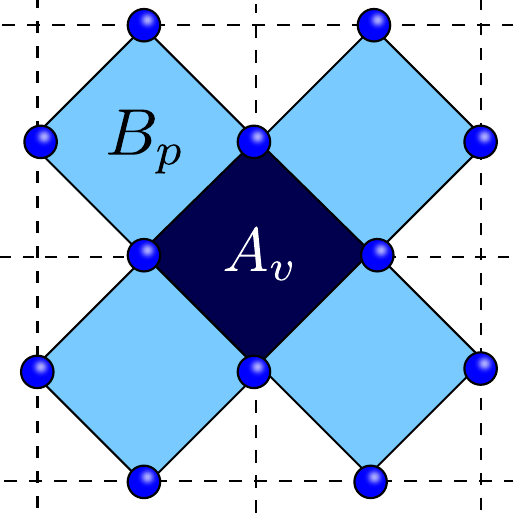}
  \end{tabular}
  
    \caption{\label{fig:lattice} Schematic of the toric code. Left: Spins are placed on the surface of a torus, with the highlighted nontrivial loop corresponding to an operation changing the logical value of the stored information. Right: The surface is structured according to a checkerboard pattern of vertex operators $A_v$ and plaquette operators $B_p$ involving the adjacent four spins of each vertex and plaquette, respectively.}
\end{figure}

The toric code \cite{Kitaev2003, kitaev2006anyons} is a paradigmatic model for a topological quantum memory. In its Hamiltonian formulation, it can be expressed as a sum of vertex and plaquette
operators
\begin{equation}
H=-\sum_{v\in\text{{vertices}}}A_{v}-\sum_{p\in\text{{plaquettes}}}B_{p}
\end{equation}
wherein the two operators are defined in terms of Paulli's acting
on spin-1/2 particles on the edges of a two-dimensional (2D) $L\times L$ periodic
lattice as shown in Fig.~\ref{fig:lattice}. The vertex and plaquette terms are given by
\begin{equation}
A_{v}=\prod_{i\in v}\sigma_{i}^{x}\quad B_{p}=\prod_{i\in p}\sigma_{i}^{z},
\end{equation}
respectively. Owing to the mutual commutativity of the operators with $H$, the
ground state(s) correspond to the state space with $A_{v}=B_{p}=1\:\forall\:v\text{{\:and}}\:p$.

\subsection{Errors in the toric code}

Local perturbations can violate the constraints on $A_v$ and $B_p$,
causing excitations above the ground state. For instance, a qubit-flip
at site-$i$ on account of local $x-$perturbation, $\sigma_{i}^{x}$
would update the two adjacent neighbouring plaquettes to
$B_{p}=-1$. These excitations at the two plaquettes lead to nontrivial
quasiparticles, known as anyons. The two anyons can
be thought of as being attached to each other by a string via the
perturbed qubit-$i$. At no further energy cost, the anyons along with
their string can diffuse further around the lattice, eventually
resulting in either a topologically trivial loop and vanishing or in
the worst case forming a topologically non-trivial loop around the
lattice. In the latter case, logical error is said to have occurred,
with the quantum system being in a different ground state orthogonal
to the initial ground state. Location of anyons are detected from
stabilizer measurements ($B_{p}$ in this case). Quantum error
correction therefore intends to pair-up these anyons forming trivial
string loops, and eliminate them altogether.

A similar description follows for another kind of anyons emerging
from $z-$perturbation or equivalently, $A_{v}=-1$. The two kinds
follow braiding statistics with one another. We focus on the uncorrelated
$x-$and $z-$perturbations scenario, where studying one kind suffice
due to the dual nature of the problem.

\subsection{\label{sec:CA}Cellular Automaton decoder}

In the following, we consider a dynamical system where errors are
continuously created and removed by different processes. The error
correction dynamics is performed by constructing a coupled
\textit{quantum-classical} system, where an open quantum system is
coupled to a cellular automaton (CA). Based on local update rules, the
classical CA guides the dynamics of the anyons in the quantum toric
code. We establish the Lindbladian dynamics of the coupled system
which involve the following processes: anyon-pair creation, anyonic
motion based on the CA field, and CA field-update. The anyon motion
and CA field-update are closely related to the operations arising in
the conventional error-correction using a CA field
\cite{Herold2015, herold2017cellular}. Below, we describe the
three processes:

1. Anyon-pair creation

\begin{equation}
E_{j}=\sigma_{j}^{x}\label{eq:E}
\end{equation}
The Paulli operator $\sigma^x$ at the $j^{\text{th}}-$spin creates
anyons on the adjacent plaquettes. We note that adding vertex
violations in the form of $\sigma^z$ errors only changes the steady
state properties quantitatively, as the vertex violations would be
updated independently. The qualitative behavior of the steady state
phase diagram will remain unchanged. To prevent the computational cost
of our simulations becoming prohibitively large. we focus on
$\sigma^x$ errors or plaquette violations only. We note that repeated
action of $E_j$ on neighboring plaquettes will also introduce diffusive
motion of the anyons.

2. Anyon motion

\begin{equation}
T_{pp'}=\sigma_{j}^{x}\left(\frac{1-B_{p}}{2}\right)w_{p'}\label{eq:T}
\end{equation}
The anyon, if present at plaquette $p$, transfers to a neighbouring
plaquette $p'$ with a probability determined by the weight factor $
w_{p'} $. Crucially, $w_{p'}$ depends on the values of the CA field
$\phi_{p'}$ at the neighboring plaquette. We choose $w_{p'}=1$ for the
neighbor $p'$ having the largest field value $\phi_{p'}$ and
$w_{p'}=0$ for all other neighbors. This means that the anyons are
moving towards the largest value of the CA field.

3. CA field-update

%\begin{eqnarray}
%F_{p}= &  & \left(\frac{1-B_{p}}{2}\right)\intop d\phi_{p}\ket{1+\tilde{\phi_{p}}-\Delta}\bra{\phi_{p}}\nonumber \\
% & + & \left(\frac{1+B_{p}}{2}\right)\intop d\phi_{p}\ket{\tilde{\phi_{p}}-\Delta}\bra{\phi_{p}}\label{eq:F}
%\end{eqnarray}
\begin{equation}
  \phi_{p}^{'}= \phi_p^{\text{avg}}+\frac{1-\mathcal{B}_p}{2} -\Delta
    \label{eq:F}
\end{equation}
Here, $\phi_{p}^{'}$ denotes the updated field value at $p$, $
\phi_p^{\text{avg}} $ is the average of the field value of the
nearest-neighbours of $p$, and $\mathcal{B}_p$ is the measurement
result obtained by measuring the operator $B_p$. The penalty term
$\Delta = \phi_p/L^2$ is introduced to regularize the growth of the
field, preventing a monotonous increase in the long-time
limit. Modulo $\Delta$, the field value at $p$ updates to average of
the field values of neighbours when anyon is absent at $p$, otherwise
it updates to average plus one.  While being local, the field-update
term spreads the local information outwards and the additional one
term attracts the anyons in the vicinity via the anyon motion process
of Eq. \eqref{eq:T}. The CA field is initialized in the state
$\phi_{p}\equiv 0$.

These processes can also be realized in a time-continuous form using a quantum master equation in Lindblad form. Specifically, the dynamics is governed by
\begin{equation}
\frac{\partial\rho}{\partial t}=\sum_{i}\left(L_{i}\rho L_{i}^{\dagger}-\frac{1}{2}\left\{ L_{i}^{\dagger}L_{i},\rho\right\} \right)\label{eq:Linbladian}
\end{equation}
where the sum runs over all the Lindbladian processes. The Lindblad
operators $L_i$ are chosen from the set $\{E_j, T_{pp'}\}$, while at
the same time the CA field is updated according to
Eq.~(\ref{eq:F}). In the following, we use $\gamma_1$, $\gamma_2$, and
$\gamma_3$ for anyon-pair creation, anyon motion, and CA field-update
rates, respectively.

Crucially, the master equation in the absence of error production,
i.e., $\gamma_1=0$, has the four ground states of the toric code as
dark states leading to $\partial_t\rho = 0$. Switching on $\gamma_1$
will perturb these states, with the central question being whether the
errors are capable to introduce nontrivial loops around the torus
corresponding to logical errors of the toric code. In particular, the
question whether the dynamics is capable to correct its own errors
establishes its own thermodynamic phase, which we will refer to as the
\emph{self-correction phase} in the following. Establishing error
correction as a thermodynamic phase allows to use the rich machinery
of quantum many-body physics to analyze its properties and its
disappearance in the form of a phase transition.

Finally, we note that the master equation can be unravelled in the
anyon basis onto a set of pure states
$\ket{\{B_p\},\{A_v=1\},\lambda}$, with $\lambda$ being a topological
quantum number \cite{Jamadagni2018}. Since these states do not contain
any superposition in this basis, the dynamics can be simulated
efficiently, allowing to reach relatively large system sizes.

\section{\label{sec:Results}Numerical results}

\subsection{Self-correction transition}

\begin{figure}
\includegraphics[width=1\columnwidth]{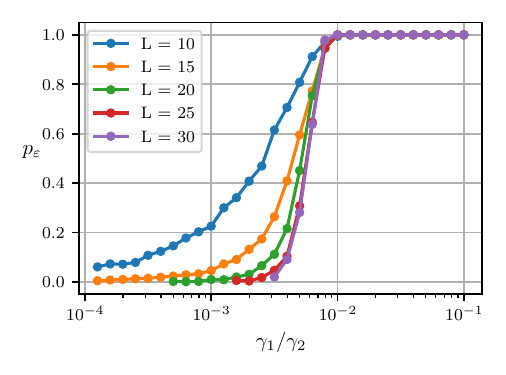}\caption{\label{fig:logical_error}
  Logical error probability $p_\varepsilon$ in the steady state
  depending on the error rate $\gamma_1$ for different system sizes
  $L\times L$. The field-update rate is taken as $\gamma_3 =
  10\,\gamma_2$. Each data point is an average obtained from
  $10^{3}-$simulation runs. The simulation points to a critical error
  rate of about $\gamma_{1}=10^{-2}\,\gamma_2$, below which logical
  errors are increasingly suppressed with system size.}
\end{figure}
Starting from the toric code ground state and with all the CA field
initialized to zero, we simulate the dynamics from
Eq. \eqref{eq:Linbladian}. The most interesting regime is
characterized by the inequalities
$\gamma_{1}\ll\gamma_{2}\ll\gamma_{3}$, which leaves the possibility
for successful suppression of errors in the long-time limit.

A central challenge in the simulation of such open systems is that we
are interested both in the long-time limit and the thermodynamic limit
at the same time, which generically do not commute with each other
\cite{Weimer2021}. To address this, we identify the timescale on which
logical errors start to appear. Under diffusive dynamics of the
anyons, this will happen on a timescale of $t\sim
L^{4} /\gamma_{1}$, which indicates how the
simulation time has to be adapted if the system size is changed to
ensure that the system has converged to its steady state. We have
numerically checked the validity of this scaling. The results are then
obtained from averaging over $10^{3}$ trajectories.

\begin{figure}
\includegraphics[width=0.9\columnwidth]{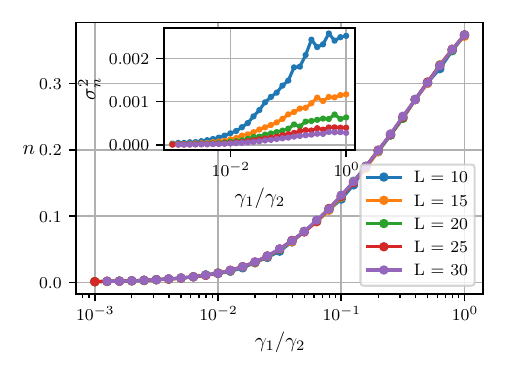}

\caption{\label{fig:anyons_count}Mean anyon density $n$ for various system
  sizes $L\times L$ as a function of the error rate $\gamma_{1}$. The inset
  depicts the variance of the anyon density $\sigma^2_n$. The variance decreases with increasing system size.}
\end{figure}
In Fig.~\ref{fig:logical_error}, we show the logical error probability
obtained for various anyon-pair creation rates. We fix $\gamma_{3}=10 \gamma_2$. We notice that for
$\gamma_{1}<10^{-2}\,\gamma_2$, the probability of a logical error falls rapidly, and this suppression persists for longer times as the system size increases. For $\gamma_{1}>10^{-2}\,\gamma_2$, the logical errors occur with
certainty in the long-time limit. As the system size scales up, we clearly observe the critical error rate converging to approximately $\gamma_{1}=10^{-2}\,\gamma_2$. This establishes the existence of a
phase transition between a phase where self-correction of errors is
possible and a second trivial phase where logical errors prevent
self-correction. Strikingly, we observe this feature even though our
CA field is only two-dimensional, which is not sufficient to obtain a
finite threshold in the conventional error correction setting \cite{Herold2015}. From
this we conclude that the time-continuous nature of the Lindblad
dynamics further stabilizes the error correction properties of the
toric code \cite{dunnweber2026quantum}.

Below the critical rate, the anyons are still sparse and the CA
generates a field that pairs them correctly without having to cross
the boundary. Interestingly, we find that the density of anyons is a
smooth function across the critical point, see
Fig.~\ref{fig:anyons_count}, which does not show significant
finite-size corrections. However, as shown in the inset, this is not
true for the variance, which strongly decreases under
finite-size scaling, although it appears to reach a finite value in the
trivial phase.

\subsection{Relationship to topological order}

Let us now turn to the question how the self-correction transition is
related to the topological order of the system. For this, we turn to
an operational definition of topological order \cite{Jamadagni2022},
which can also be readily applied for open quantum systems. Specifically,
we compute the depth of a classical error correction circuit to
correct the system back into one of the ground states of the toric
code. The concrete implementation associates with each measured error
a walker that explores its surroundings to fuse with other errors
\cite{Wootton2015}. Then, if the depth is sufficiently short, we
conclude that the system is in its topologically ordered phase.

\begin{figure}
\includegraphics[width=1\columnwidth]{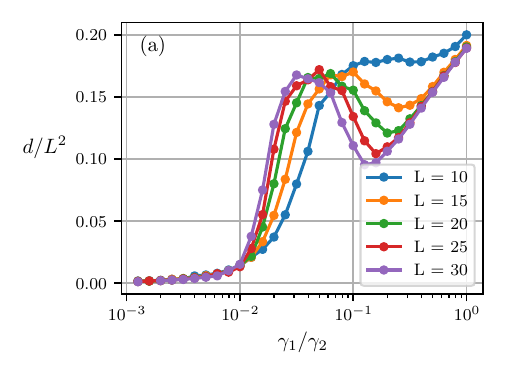}

\includegraphics[width=1\columnwidth]{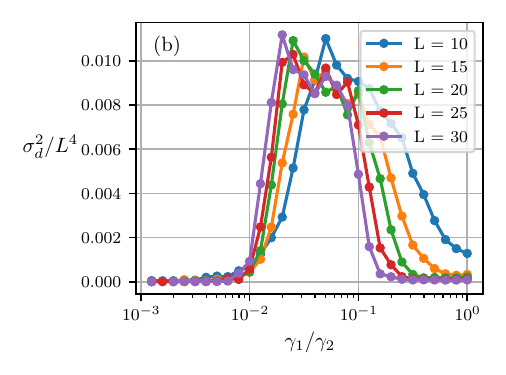}\caption{\label{fig:circuit_depth} (a) Mean circuit depth $d$ normalized to the system size $L\times L$ as a function of the error rate $\gamma_1$ with fixed $\gamma_3 = 10\,\gamma_2$. Above the self-correction transition, the normalized circuit depth shows a modified scaling with system size. (b)~Variance of the circuit depth $\sigma_d^2$ normalized to the system size. Below the transition, the normalized variance vanishes in the thermodynamic limit, while it reaches a finite value after the transition.}
\end{figure}

In Fig. \ref{fig:circuit_depth}a, we show the mean circuit depth as a
function of $\gamma_1$, normalized to the system size $L^2$. As
expected, we observe a qualitative change in the finite-size scaling
behavior when crossing the self-correction transition. At small
$\gamma_1$ before the transition, there is a data collapse of all
system sizes onto a single line. After the transition, there is a
region of superlinear growth of the circuit depth. Remarkably, the
normalized circuit depth being constant in the self-correcting phase
implies that the unnormalized circuit depth increases with the system
size. This is a different situation compared to many other examples,
where the circuit depth in the topological phase does not scale with
the system size
\cite{Jamadagni2022, Jamadagni_2022, Jamadagni_2023, Fraatz_2026}. Nevertheless,
the growth with system size is not sufficient to cause logical errors,
as the mean circuit depth only reaches a few percent of the total
system size. This suggests that fluctuations of the circuit depth have
to be sufficiently small, so they cannot induce logical errors in a
significant fraction of the observed realizations. Indeed, as shown in
Fig.~\ref{fig:circuit_depth}b, the variance of the normalized circuit
depth vanishes in the self-correcting phase, while it reaches a finite
value following the phase transition. From this we conclude that the
variance of the circuit depth is a better measure to determine the
phase transition.

Turning to the operational definition also has a practical
advantage. In the previous section, we have computed the logical error
probability for the initial state being one of the toric code ground
states. However, it is not possible to use a trivial initial state in
this approach, as the value of the logical bit is not defined in this
case. In contrast to this, the operational definition can be applied
for arbitrary initial states. We have computed the circuit depth using the topologically trivial maximally mixed state as the initial state and obtain results that are indistinguishable from those presented in Fig.~\ref{fig:circuit_depth}.

\subsection{Steady-state phase diagram}

\begin{figure}
  \includegraphics[width=1\columnwidth]{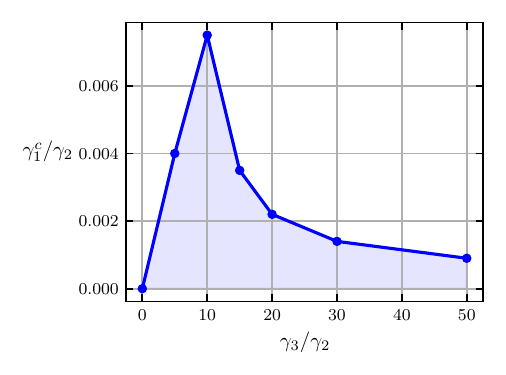}\caption{\label{fig:phase_diag}
    Phase diagram depicting the critical error rate $\gamma_1^c$
    depending on the CA field-update rate $\gamma_3$. The shaded region
    indicates the self-correcting phase in which errors can
    successfully be corrected. This suggests an optimal rate of CA field-update beyond which the critical error rate starts
    declining.}
\end{figure}
We now obtain the entire phase diagram for different rates of the
processes in the master equation using the operational
definition. Using the error correction rate $\gamma_2$ as a fixed
reference, we compute the critical rate $\gamma_1$ for the
self-correction transition for different choices of the field-update
rate $\gamma_3$. The shaded region in Fig.~\ref{fig:phase_diag}
corresponds to the self-correcting phase. Remarkably, we do not
observe a monotonous behavior of the phase transition line, but rather
a pronounced peak around an optimal field-update rate. While it can be
readily understood that a slow field-update rate can be insufficient
to provide the information for the anyons to move in the correct
direction, the behavior for large $\gamma_3$ deserves further
explanation. We attribute the decrease in the critical $\gamma_1$ to
the fact that for large $\gamma_3$, the local field will also contain
information from more distant anyons. However, the key notion behind
error correction in the toric code is to fuse all emerging anyon pairs
in the same way they were originally created. Generically, anyons that
are further apart are more likely to stem from a different anyon pair,
making it unfavorable to move towards them and thus reducing the
critical $\gamma_1$.

\section{\label{sec:Conclusion}Conclusion}

In this work, we have analyzed a quantum master equation derived from
a CA decoder for topological quantum error correction. Despite the
update rules being strictly local, we find the existence of a
self-correcting phase in the steady state of the dynamics. The
decoding emerges from local, non-equilibrium
dynamics of both the quantum system under consideration and the
classical CA field. Our numerical simulations establish a
self-correction threshold when the ratio between error generation and
error correction rates is about $0.008$.

From a practical implementation standpoint, our approach offers
significant advantages for scalable quantum computing architectures.
Unlike matching-based decoders, which often face a latency bottleneck
as system sizes grow, the CA approach allows for $\mathcal{O}(1)$
complexity per field-update. This locality permits the integration
of decoding logic directly into the qubit control lines \cite{caune2024FPGA}, significantly
reducing the bandwidth requirements between the quantum plane and
the classical control stack.

\bibliographystyle{apsrev4-2}
\bibliography{ref}

@Article{Weimer2021,
  title = {Simulation methods for open quantum many-body systems},
  author = {Weimer, Hendrik and Kshetrimayum, Augustine and Or\'us, Rom\'an},
  journal = {Rev. Mod. Phys.},
  volume = {93},
  issue = {1},
  pages = {015008},
  numpages = {24},
  year = {2021},
  month = {Mar},
  publisher = {American Physical Society},
  doi = {10.1103/RevModPhys.93.015008},
  url = {https://link.aps.org/doi/10.1103/RevModPhys.93.015008}
}

@Article{Wootton2015,
  Title                    = {A Simple Decoder for Topological Codes},
  Author                   = {Wootton, James},
  Journal                  = {Entropy},
  Year                     = {2015},
  Number                   = {4},
  Pages                    = {1946--1957},
  Volume                   = {17},
  Doi                      = {10.3390/e17041946},
  ISSN                     = {1099-4300},
  Url                      = {http://www.mdpi.com/1099-4300/17/4/1946}
}

@Article{Jamadagni2018,
  Title                    = {Robustness of topological order in the toric code with open boundaries},
  Author                   = {Jamadagni, Amit and Weimer, Hendrik and Bhattacharyya, Arpan},
  Journal                  = {Phys. Rev. B},
  Year                     = {2018},

  Month                    = {Dec},
  Pages                    = {235147},
  Volume                   = {98},

  Doi                      = {10.1103/PhysRevB.98.235147},
  Issue                    = {23},
  Numpages                 = {11},
  Publisher                = {American Physical Society},
  Url                      = {https://link.aps.org/doi/10.1103/PhysRevB.98.235147}
}

@article{Sieberer2025,
  title = {Universality in driven open quantum matter},
  author = {Sieberer, Lukas M. and Buchhold, Michael and Marino, Jamir and Diehl, Sebastian},
  journal = {Rev. Mod. Phys.},
  volume = {97},
  issue = {2},
  pages = {025004},
  numpages = {66},
  year = {2025},
  month = {Jun},
  publisher = {American Physical Society},
  doi = {10.1103/RevModPhys.97.025004},
  url = {https://link.aps.org/doi/10.1103/RevModPhys.97.025004}
}

@Article{Mink2023,
	title={{Collective radiative interactions in the discrete truncated Wigner approximation}},
	author={Christopher D. Mink and Michael Fleischhauer},
	journal={SciPost Phys.},
	volume={15},
	pages={233},
	year={2023},
	publisher={SciPost},
	doi={10.21468/SciPostPhys.15.6.233},
	url={https://scipost.org/10.21468/SciPostPhys.15.6.233},
}

@article{Singh2022,
	title = {Driven-Dissipative Criticality within the Discrete Truncated Wigner Approximation},
	author = {Singh, Vijay Pal and Weimer, Hendrik},
	journal = {Phys. Rev. Lett.},
	volume = {128},
	issue = {20},
	pages = {200602},
	numpages = {7},
	year = {2022},
	month = {May},
	publisher = {American Physical Society},
	doi = {10.1103/PhysRevLett.128.200602},
	url = {https://link.aps.org/doi/10.1103/PhysRevLett.128.200602}
}

@article{Carollo2022,
  title = {Nonequilibrium Dark Space Phase Transition},
  author = {Carollo, Federico and Lesanovsky, Igor},
  journal = {Phys. Rev. Lett.},
  volume = {128},
  issue = {4},
  pages = {040603},
  numpages = {7},
  year = {2022},
  month = {Jan},
  publisher = {American Physical Society},
  doi = {10.1103/PhysRevLett.128.040603},
  url = {https://link.aps.org/doi/10.1103/PhysRevLett.128.040603}
}

@Article{	  Panas2019,
  title		= {Density-wave steady-state phase of dissipative ultracold
		  fermions with nearest-neighbor interactions},
  author	= {Panas, Jaromir and Pasek, Michael and Dhar, Arya and Qin,
		  Tao and Gei\ss{}ler, Andreas and Hafez-Torbati, Mohsen and
		  Sorantin, Max E. and Titvinidze, Irakli and Hofstetter,
		  Walter},
  journal	= {Phys. Rev. B},
  volume	= {99},
  issue		= {11},
  pages		= {115125},
  numpages	= {14},
  year		= {2019},
  month		= {Mar},
  publisher	= {American Physical Society},
  doi		= {10.1103/PhysRevB.99.115125},
  url		= {https://link.aps.org/doi/10.1103/PhysRevB.99.115125}
}

@article{Carollo2019,
  title = {Critical Behavior of the Quantum Contact Process in One Dimension},
  author = {Carollo, Federico and Gillman, Edward and Weimer, Hendrik and Lesanovsky, Igor},
  journal = {Phys. Rev. Lett.},
  volume = {123},
  issue = {10},
  pages = {100604},
  numpages = {6},
  year = {2019},
  month = {Sep},
  publisher = {American Physical Society},
  doi = {10.1103/PhysRevLett.123.100604},
  url = {https://link.aps.org/doi/10.1103/PhysRevLett.123.100604}
}

@Article{Nagy2018,
  title = {Driven-dissipative quantum Monte Carlo method for open quantum systems},
  author = {Nagy, Alexandra and Savona, Vincenzo},
  journal = {Phys. Rev. A},
  volume = {97},
  issue = {5},
  pages = {052129},
  numpages = {9},
  year = {2018},
  month = {May},
  publisher = {American Physical Society},
  doi = {10.1103/PhysRevA.97.052129},
  url = {https://link.aps.org/doi/10.1103/PhysRevA.97.052129}
}

@article{Buchhold2017,
  title = {Nonequilibrium effective field theory for absorbing state phase transitions in driven open quantum spin systems},
  author = {Buchhold, Michael and Everest, Benjamin and Marcuzzi, Matteo and Lesanovsky, Igor and Diehl, Sebastian},
  journal = {Phys. Rev. B},
  volume = {95},
  issue = {1},
  pages = {014308},
  numpages = {31},
  year = {2017},
  month = {Jan},
  publisher = {American Physical Society},
  doi = {10.1103/PhysRevB.95.014308},
  url = {https://link.aps.org/doi/10.1103/PhysRevB.95.014308}
}

@article{Rota2017,
  title = {Critical behavior of dissipative two-dimensional spin lattices},
  author = {Rota, R. and Storme, F. and Bartolo, N. and Fazio, R. and Ciuti, C.},
  journal = {Phys. Rev. B},
  volume = {95},
  issue = {13},
  pages = {134431},
  numpages = {5},
  year = {2017},
  month = {Apr},
  publisher = {American Physical Society},
  doi = {10.1103/PhysRevB.95.134431},
  url = {https://link.aps.org/doi/10.1103/PhysRevB.95.134431}
}

@article{Marino2016,
  title = {Driven Markovian Quantum Criticality},
  author = {Marino, Jamir and Diehl, Sebastian},
  journal = {Phys. Rev. Lett.},
  volume = {116},
  issue = {7},
  pages = {070407},
  numpages = {6},
  year = {2016},
  month = {Feb},
  publisher = {American Physical Society},
  doi = {10.1103/PhysRevLett.116.070407},
  url = {https://link.aps.org/doi/10.1103/PhysRevLett.116.070407}
}

@Article{Jin2016,
  Title                    = {Cluster Mean-Field Approach to the Steady-State Phase
 Diagram of Dissipative Spin Systems},
  Author                   = {Jin, Jiasen and Biella, Alberto and Viyuela, Oscar and
 Mazza, Leonardo and Keeling, Jonathan and Fazio, Rosario
 and Rossini, Davide},
  Journal                  = {Phys. Rev. X},
  Year                     = {2016},

  Month                    = {Jul},
  Pages                    = {031011},
  Volume                   = {6},

  Doi                      = {10.1103/PhysRevX.6.031011},
  Issue                    = {3},
  Numpages                 = {18},
  Publisher                = {American Physical Society},
  Url                      = {https://link.aps.org/doi/10.1103/PhysRevX.6.031011}
}

@Article{Maghrebi2016,
  Title                    = {Nonequilibrium many-body steady states via Keldysh
 formalism},
  Author                   = {Maghrebi, Mohammad F. and Gorshkov, Alexey V.},
  Journal                  = {Phys. Rev. B},
  Year                     = {2016},

  Month                    = {Jan},
  Pages                    = {014307},
  Volume                   = {93},

  Doi                      = {10.1103/PhysRevB.93.014307},
  Issue                    = {1},
  Numpages                 = {15},
  Publisher                = {American Physical Society},
  Url                      = {http://link.aps.org/doi/10.1103/PhysRevB.93.014307}
}

@article{Marcuzzi2016,
  title = {Absorbing State Phase Transition with Competing Quantum and Classical Fluctuations},
  author = {Marcuzzi, Matteo and Buchhold, Michael and Diehl, Sebastian and Lesanovsky, Igor},
  journal = {Phys. Rev. Lett.},
  volume = {116},
  issue = {24},
  pages = {245701},
  numpages = {7},
  year = {2016},
  month = {Jun},
  publisher = {American Physical Society},
  doi = {10.1103/PhysRevLett.116.245701},
  url = {https://link.aps.org/doi/10.1103/PhysRevLett.116.245701}
}

@Article{Cai2013,
  Title                    = {Algebraic versus Exponential Decoherence in Dissipative
 Many-Particle Systems},
  Author                   = {Cai, Zi and Barthel, Thomas},
  Journal                  = {Phys. Rev. Lett.},
  Year                     = {2013},

  Month                    = {Oct},
  Pages                    = {150403},
  Volume                   = {111},

  Doi                      = {10.1103/PhysRevLett.111.150403},
  Issue                    = {15},
  Numpages                 = {5},
  Publisher                = {American Physical Society},
  Url                      = {http://link.aps.org/doi/10.1103/PhysRevLett.111.150403}
}

@Article{Weimer2015,
  Title                    = {Variational Principle for Steady States of Dissipative
 Quantum Many-Body Systems},
  Author                   = {Weimer, Hendrik},
  Journal                  = {Phys. Rev. Lett.},
  Year                     = {2015},

  Month                    = {Jan},
  Pages                    = {040402},
  Volume                   = {114},

  Doi                      = {10.1103/PhysRevLett.114.040402},
  Issue                    = {4},
  Numpages                 = {6},
  Publisher                = {American Physical Society}
}

@Article{Lee2013,
  Title                    = {Unconventional Magnetism via Optical Pumping of
 Interacting Spin Systems},
  Author                   = {Lee, Tony E. and Gopalakrishnan, Sarang and Lukin, Mikhail
 D.},
  Journal                  = {Phys. Rev. Lett.},
  Year                     = {2013},

  Month                    = {Jun},
  Pages                    = {257204},
  Volume                   = {110},

  Doi                      = {10.1103/PhysRevLett.110.257204},
  Issue                    = {25},
  Numpages                 = {5},
  Publisher                = {American Physical Society},
  Url                      = {http://link.aps.org/doi/10.1103/PhysRevLett.110.257204}
}

@Article{Torre2013,
  Title                    = {Keldysh approach for nonequilibrium phase transitions in
 quantum optics: Beyond the Dicke model in optical
 cavities},
  Author                   = {Torre, Emanuele G. Dalla and Diehl, Sebastian and Lukin,
 Mikhail D. and Sachdev, Subir and Strack, Philipp},
  Journal                  = {Phys. Rev. A},
  Year                     = {2013},

  Month                    = {Feb},
  Pages                    = {023831},
  Volume                   = {87},

  Doi                      = {10.1103/PhysRevA.87.023831},
  Issue                    = {2},
  Numpages                 = {20},
  Publisher                = {American Physical Society},
  Url                      = {http://link.aps.org/doi/10.1103/PhysRevA.87.023831}
}

@Article{Sieberer2013,
  Title                    = {Dynamical Critical Phenomena in Driven-Dissipative
 Systems},
  Author                   = {Sieberer, L. M. and Huber, S. D. and Altman, E. and Diehl,
 S.},
  Journal                  = {Phys. Rev. Lett.},
  Year                     = {2013},

  Month                    = {May},
  Pages                    = {195301},
  Volume                   = {110},

  Doi                      = {10.1103/PhysRevLett.110.195301},
  Issue                    = {19},
  Numpages                 = {5},
  Publisher                = {American Physical Society},
  Url                      = {http://link.aps.org/doi/10.1103/PhysRevLett.110.195301}
}

@Article{Honing2012,
  Title                    = {Critical exponents of steady-state phase transitions in
 fermionic lattice models},
  Author                   = {H\"oning, M. and Moos, M. and Fleischhauer, M.},
  Journal                  = {Phys. Rev. A},
  Year                     = {2012},

  Month                    = {Jul},
  Pages                    = {013606},
  Volume                   = {86},

  Doi                      = {10.1103/PhysRevA.86.013606},
  Issue                    = {1},
  Numpages                 = {7},
  Publisher                = {American Physical Society},
  Url                      = {http://link.aps.org/doi/10.1103/PhysRevA.86.013606}
}

@Article{Lee2011,
  Title                    = {Antiferromagnetic phase transition in a nonequilibrium
 lattice of Rydberg atoms},
  Author                   = {Lee, Tony E. and H\"affner, H. and Cross, M. C.},
  Journal                  = {Phys. Rev. A},
  Year                     = {2011},

  Month                    = {Sep},
  Pages                    = {031402(R)},
  Volume                   = {84},

  Doi                      = {10.1103/PhysRevA.84.031402},
  Issue                    = {3},
  Numpages                 = {4},
  Publisher                = {American Physical Society},
  Url                      = {http://link.aps.org/doi/10.1103/PhysRevA.84.031402}
}

@Article{Hartmann2010,
  Title                    = {Polariton Crystallization in Driven Arrays of Lossy
 Nonlinear Resonators},
  Author                   = {Hartmann, Michael J.},
  Journal                  = {Phys. Rev. Lett.},
  Year                     = {2010},

  Month                    = {Mar},
  Pages                    = {113601},
  Volume                   = {104},

  Doi                      = {10.1103/PhysRevLett.104.113601},
  Issue                    = {11},
  Numpages                 = {4},
  Publisher                = {American Physical Society},
  Url                      = {http://link.aps.org/doi/10.1103/PhysRevLett.104.113601}
}

@Article{Tomadin2011,
  Title                    = {Nonequilibrium phase diagram of a driven and dissipative
 many-body system},
  Author                   = {Tomadin, Andrea and Diehl, Sebastian and Zoller, Peter},
  Journal                  = {Phys. Rev. A},
  Year                     = {2011},

  Month                    = {Jan},
  Pages                    = {013611},
  Volume                   = {83},

  Doi                      = {10.1103/PhysRevA.83.013611},
  Issue                    = {1},
  Numpages                 = {20},
  Publisher                = {American Physical Society},
  Url                      = {http://link.aps.org/doi/10.1103/PhysRevA.83.013611}
}

@Article{Diehl2010a,
  Title                    = {Dynamical Phase Transitions and Instabilities in Open
 Atomic Many-Body Systems},
  Author                   = {Diehl, Sebastian and Tomadin, Andrea and Micheli, Andrea
 and Fazio, Rosario and Zoller, Peter},
  Journal                  = {Phys. Rev. Lett.},
  Year                     = {2010},

  Month                    = {Jul},
  Pages                    = {015702},
  Volume                   = {105},

  Doi                      = {10.1103/PhysRevLett.105.015702},
  Issue                    = {1},
  Numpages                 = {4},
  Publisher                = {American Physical Society},
  Url                      = {http://link.aps.org/doi/10.1103/PhysRevLett.105.015702}
}

@Article{Nagy2010,
  Title                    = {Dicke-Model Phase Transition in the Quantum Motion of a
 Bose-Einstein Condensate in an Optical Cavity},
  Author                   = {Nagy, D. and K\'onya, G. and Szirmai, G. and Domokos, P.},
  Journal                  = {Phys. Rev. Lett.},
  Year                     = {2010},

  Month                    = {Apr},
  Pages                    = {130401},
  Volume                   = {104},

  Doi                      = {10.1103/PhysRevLett.104.130401},
  Issue                    = {13},
  Numpages                 = {4},
  Publisher                = {American Physical Society},
  Url                      = {http://link.aps.org/doi/10.1103/PhysRevLett.104.130401}
}

@article{Kitaev2003,
  Title                    = {Fault-tolerant quantum computation by anyons},
  Author                   = {A. Kitaev},
  Journal                  = {Annals of Physics},
  Year                     = {2003},
  Pages                    = {2--30},
  Volume                   = {303},
  doi = {https://doi.org/10.1016/S0003-4916(02)00018-0}
}

@article{Raghunandan2020,
       author = {{Raghunandan}, Meghana and {Wolf}, Fabian and {Ospelkaus}, Christian and {Schmidt}, Piet O. and {Weimer}, Hendrik},
        title = "{Initialization of quantum simulators by sympathetic cooling}",
      journal = {Science Advances},
         year = 2020,
        month = mar,
       volume = {6},
       number = {10},
        pages = {eaaw9268},
          doi = {10.1126/sciadv.aaw9268}
}

@article{Kshetrimayum2017,
  Title                    = {{A simple tensor network algorithm for two-dimensional
 steady states}},
  Author                   = {{Kshetrimayum}, A. and {Weimer}, H. and {Or{\'u}s}, R.},
  Journal                  = {Nature Commun.},
  Year                     = {2017},

  Month                    = nov,
  Pages                    = {1291},
  Volume                   = {8},
  Doi                      = {10.1038/s41467-017-01511-6}
}

@article{DiVincenzo2000,
  Title                    = {The Physical Implementation of Quantum Computation},
  Author                   = {DiVincenzo, David P.},
  Journal                  = {Fortschr. Phys.},
  Year                     = {2000},
  Pages                    = {771-783},
  Volume                   = {48},
  doi = {https://doi.org/10.1002/1521-3978(200009)48:9/11<771::AID-PROP771>3.0.CO;2-E}
}

@article{Jamadagni_2023,
  title = {Learning of error statistics for the detection of quantum phases},
  author = {Jamadagni, Amit and Kazemi, Javad and Weimer, Hendrik},
  journal = {Phys. Rev. B},
  volume = {107},
  issue = {7},
  pages = {075146},
  numpages = {9},
  year = {2023},
  month = {Feb},
  publisher = {American Physical Society},
  doi = {10.1103/PhysRevB.107.075146},
  url = {https://link.aps.org/doi/10.1103/PhysRevB.107.075146}
}

@article{Jamadagni_2022,
  title = {Error-correction properties of an interacting topological insulator},
  author = {Jamadagni, Amit and Weimer, Hendrik},
  journal = {Phys. Rev. B},
  volume = {106},
  issue = {11},
  pages = {115133},
  numpages = {6},
  year = {2022},
  month = {Sep},
  publisher = {American Physical Society},
  doi = {10.1103/PhysRevB.106.115133},
  url = {https://link.aps.org/doi/10.1103/PhysRevB.106.115133}
}

@article{Fraatz_2026,
doi = {10.1088/1367-2630/ae3598},
url = {https://doi.org/10.1088/1367-2630/ae3598},
year = {2026},
month = {jan},
publisher = {IOP Publishing},
volume = {28},
number = {1},
pages = {014512},
author = {Fraatz, Louis and Jamadagni, Amit and Weimer, Hendrik},
title = {Efficient computation of topological order},
journal = {New Journal of Physics},
}

@book{nielsen2010quantum,
  title={Quantum computation and quantum information},
  author={Nielsen, Michael A and Chuang, Isaac L},
  year={2010},
  publisher={Cambridge University Press},
  doi = {https://doi.org/10.1017/CBO9780511976667}
}

@article{preskill1998reliable,
  title={Reliable quantum computers},
  author={Preskill, John},
  journal={Proceedings of the Royal Society of London. Series A},
  volume={454},
  number={1969},
  pages={385--410},
  year={1998},
  publisher={The Royal Society},
  doi = {https://doi.org/10.1098/rspa.1998.0167}
}

@article{terhal2015quantum,
  title={Quantum error correction for quantum memories},
  author={Terhal, Barbara M},
  journal={Reviews of Modern Physics},
  volume={87},
  number={2},
  pages={307--346},
  year={2015},
  publisher={APS},
  doi = {10.1103/RevModPhys.87.307}
}

@book{gottesman1997stabilizer,
  title={Stabilizer codes and quantum error correction},
  author={Gottesman, Daniel},
  year={1997},
  publisher={California Institute of Technology},
  doi={https://doi.org/10.48550/arXiv.quant-ph/9705052}
}

@article{Weimer2010,
  Title                    = {A Rydberg quantum simulator},
  Author                   = {{Weimer}, H. and {M{\"u}ller}, M. and {Lesanovsky}, I. and
 {Zoller}, P. and {B{\"u}chler}, H.~P.},
  Journal                  = {Nature Phys.},
  Year                     = {2010},
  Pages                    = {382--388},
  Volume                   = {6},

  Doi                      = {10.1038/NPHYS1614}
}

@article{Diehl2008,
	author = {Diehl, S. and Micheli, A. and Kantian, A. and Kraus, B. and B{\"u}chler, H. P. and Zoller, P.},
	da = {2008/11/01},
	doi = {10.1038/nphys1073},
	id = {Diehl2008},
	isbn = {1745-2481},
	journal = {Nature Physics},
	number = {11},
	pages = {878--883},
	title = {Quantum states and phases in driven open quantum systems with cold atoms},
	ty = {JOUR},
	url = {https://doi.org/10.1038/nphys1073},
	volume = {4},
	year = {2008}
}

@article{Verstraete2009,
  Title                    = {Quantum computation and quantum-state engineering driven
 by dissipation},
  Author                   = {Verstraete, Frank and Wolf, Michael M. and Ignacio Cirac,
 J.},
  Journal                  = {Nature Phys.},
  Year                     = {2009},

  Month                    = {Sep},
  Number                   = {9},
  Pages                    = {633--636},
  Volume                   = {5},

  Doi                      = {10.1038/nphys1342},
  Publisher                = {Nature Publishing Group}
}

@article{wang2003confinement,
  title={Confinement-Higgs transition in a disordered gauge theory and the accuracy threshold for quantum memory},
  author={Wang, Chenyang and Harrington, Jim and Preskill, John},
  journal={Annals of Physics},
  volume={303},
  number={1},
  pages={31--58},
  year={2003},
  publisher={Elsevier},
  doi = {https://doi.org/10.1016/S0003-4916(02)00019-2}
}

@article{Vodola2022fundamental,
  doi = {10.22331/q-2022-01-05-618},
  url = {https://doi.org/10.22331/q-2022-01-05-618},
  title = {Fundamental thresholds of realistic quantum error correction circuits from classical spin models},
  author = {Vodola, Davide and Rispler, Manuel and Kim, Seyong and M{\"{u}}ller, Markus},
  journal = {{Quantum}},
  issn = {2521-327X},
  publisher = {{Verein zur F{\"{o}}rderung des Open Access Publizierens in den Quantenwissenschaften}},
  volume = {6},
  pages = {618},
  month = jan,
  year = {2022}
}

@article{Kessler2012,
  title={Dissipative phase transition in a central spin system},
  author={Kessler, Eric M and Giedke, Geza and Imamoglu, Atac and Yelin, Susanne F and Lukin, Mikhail D and Cirac, J Ignacio},
  journal={Physical Review A},
  volume={86},
  number={1},
  pages={012116},
  year={2012},
  publisher={APS},
  doi = {https://doi.org/10.1103/PhysRevA.86.012116}
}

@article{minganti2018spectral,
  title={Spectral theory of Liouvillians for dissipative phase transitions},
  author={Minganti, Fabrizio and Biella, Alberto and Bartolo, Nicola and Ciuti, Cristiano},
  journal={Physical Review A},
  volume={98},
  number={4},
  pages={042118},
  year={2018},
  publisher={APS},
  doi = {https://doi.org/10.1103/PhysRevA.98.042118}
}

@article{bardyn2013topology,
  title={Topology by dissipation},
  author={Bardyn, Charles-Edouard and Baranov, Mikhail A and Kraus, Christina V and Rico, Enrique and {\.I}mamo{\u{g}}lu, A and Zoller, Peter and Diehl, Sebastian},
  journal={New Journal of Physics},
  volume={15},
  number={8},
  pages={085001},
  year={2013},
  publisher={IOP Publishing},
  doi={10.1088/1367-2630/15/8/085001}
}

@article{dunnweber2026quantum,
  title={Quantum Memory and Autonomous Computation in Two Dimensions},
  author={D{\"u}nnweber, Gesa and Styliaris, Georgios and Trivedi, Rahul},
  journal={arXiv:2601.20818},
  year={2026},
  url={https://arxiv.org/abs/2601.20818}
}

\end{document}